\documentclass[longauth]{aa} % for the long lists of affiliations 
\usepackage{graphicx}
\usepackage{txfonts}
\usepackage{amssymb}
\usepackage{amsmath}
\usepackage{graphicx}
\usepackage{float}
\usepackage[colorlinks=true, allcolors=blue]{hyperref}
\usepackage{booktabs}
\usepackage{amsfonts}     % matemaattiset merkit
\usepackage[T1]{fontenc} %font encoding
\usepackage[utf8]{inputenc} %unicode
\usepackage{multicol}
\usepackage{xcolor}
\usepackage{soul}
\usepackage{rotating}
\usepackage{comment}
\usepackage{multirow}
\usepackage{longtable}
\usepackage{lscape}
\usepackage{color}
\usepackage{url}

\usepackage{orcidlink}
\usepackage{academicons}

\linenumbers

\begin{document}

\title{Determining the origin of the X-ray emission in blazars through multiwavelength polarization}

\author{
Ioannis Liodakis, \inst{\ref{AstroCrete},\ref{MPI_Bonn},\ref{NASA_Alabama}} \thanks{\href{mailto:liodakis@ia.forth.gr}{liodakis@ia.forth.gr}} \orcid{0000-0001-9200-4006} %liodakis@ia.forth.gr
Haocheng Zhang, \inst{\ref{Maryland},\ref{NASA_goddard}} \orcid{0000-0001-9826-1759}%haocheng.zhang@nasa.gov
Stella Boula,\inst{\ref{INAF_Merate}}\orcid{0000-0001-7905-692+8}
Riccardo Middei, \inst{\ref{SSDC_Rome},\ref{INAF_Rome_Obs}} \orcid{0000-0001-9815-9092} %riccardo.middei@ssdc.asi.it
Jorge Otero-Santos, \inst{\ref{InstAstro_Granada},\ref{INFN-Padova}} %\orcid{} %joteros@iaa.es
Dmitry Blinov, \inst{\ref{AstroCrete},\ref{AstroCrete2}} %dmitriy.blinov@gmail.com
Iv\'{a}n Agudo, \inst{\ref{InstAstro_Granada}} \orcid{0000-0002-3777-6182} %iagudo@iaa.es
Markus B\"ottcher, \inst{\ref{Northwest}} \orcid{0000-0002-8434-5692}%-----------------------------------------------------------------
Chien-Ting Chen, \inst{\ref{USRA_Alabama}} \orcid{0000-0002-4945-5079} %chien-ting.chen@nasa.gov
Steven R. Ehlert, \inst{\ref{NASA_Alabama}} \orcid{0000-0003-4420-2838} %steven.r.ehlert@nasa.gov
Svetlana G. Jorstad, \inst{\ref{BostonUni},\ref{StPetersburg}} \orcid{0000-0001-6158-1708} 
Philip Kaaret, \inst{\ref{NASA_Alabama}} \orcid{0000-0002-3638-0637} %philip.kaaret@nasa.gov    
Henric Krawczynski, \inst{\ref{StLouis}} \orcid{0000-0002-1084-6507} %krawcz@wustl.edu
%Alan P. Marscher, \inst{\ref{BostonUni}} \orcid{0000-0001-7396-3332} %marscher@bu.edu
Abel L. Peirson, \inst{\ref{Stanford}} \orcid{0000-0001-6292-1911} %alpv95@alumni.stanford.edu
Roger W. Romani, \inst{\ref{Stanford}} \orcid{0000-0001-6711-3286} %rwr@astro.stanford.edu
Fabrizio Tavecchio, \inst{\ref{INAF_Merate}} \orcid{0000-0003-0256-0995} %fabrizio.tavecchio@inaf.it
Martin C. Weisskopf, \inst{\ref{NASA_Alabama}} \orcid{0000-0002-5270-4240} %martin.c.weisskopf@nasa.gov
Pouya M. Kouch, \inst{\ref{UTU},\ref{FINCA},\ref{MRO}} \orcid{0000-0002-9328-2750}
Elina Lindfors, \inst{\ref{UTU},\ref{FINCA}} \orcid{0000-0002-9155-6199} %elilin@utu.fi
Kari Nilsson, \inst{\ref{FINCA}} \orcid{0000-0002-1445-8683} %kari.nilsson@utu.fi
Callum McCall, \inst{\ref{LJMU}} %c.mccall@2017.ljmu.ac.uk
Helen E. Jermak, \inst{\ref{LJMU}} %h.e.jermak@ljmu.ac.uk
Iain A. Steele, \inst{\ref{LJMU}} %i.a.steele@ljmu.ac.uk
Ioannis Myserlis, \inst{\ref{IRAM},\ref{MPI_Bonn}} \orcid{0000-0003-3025-9497} %imyserlis@iram.es
Mark Gurwell, \inst{\ref{Harvard_Smithsonian}} \orcid{0000-0003-0685-3621} %mgurwell@cfa.harvard.edu
Garrett K. Keating, \inst{\ref{Harvard_Smithsonian}} \orcid{0000-0002-3490-146X} %garrett.keating@cfa.harvard.edu
Ramprasad Rao, \inst{\ref{Harvard_Smithsonian}} %\orcid{} %rrao@cfa.harvard.edu
Sincheol Kang, \inst{\ref{Korea_ASSI}} \orcid{0000-0002-0112-4836} %kang87@kasi.re.kr
Sang-Sung Lee, \inst{\ref{Korea_ASSI},\ref{Korea_UniSciTech}} \orcid{0000-0002-6269-594X} %sslee@kasi.re.kr
Sanghyun Kim, \inst{\ref{Korea_ASSI},\ref{Korea_UniSciTech}} \orcid{0000-0001-7556-8504} %sanghkim@kasi.re.kr
Whee Yeon Cheong, \inst{\ref{Korea_ASSI},\ref{Korea_UniSciTech}} \orcid{0009-0002-1871-5824} %wheeyeon@kasi.re.kr
Hyeon-Woo Jeong, \inst{\ref{Korea_ASSI},\ref{Korea_UniSciTech}} \orcid{0009-0005-7629-8450} %hwjeong@kasi.re.kr
Emmanouil Angelakis, \inst{\ref{Greece_UniAthens}} \orcid{0000-0001-7327-5441} %eangelakis@physics.auth.gr
Alexander Kraus, \inst{\ref{MPI_Bonn}} \orcid{0000-0002-4184-9372} %akraus@mpifr-bonn.mpg.de
Francisco Jos\'e Aceituno, \inst{\ref{InstAstro_Granada}} %\orcid{} %fja@iaa.es
Giacomo Bonnoli, \inst{\ref{INAF_Merate},\ref{InstAstro_Granada}} \orcid{0000-0003-2464-9077} %giacomo.bonnoli@inaf.it
V\'{i}ctor Casanova, \inst{\ref{InstAstro_Granada}} %\orcid{} %casanova@iaa.es
Juan Escudero, \inst{\ref{InstAstro_Granada}} \orcid{0000-0002-4131-655X} %jescudero@iaa.es
Beatriz Ag\'{i}s-Gonz\'{a}lez, \inst{\ref{AstroCrete}}  %\orcid{} %bagis@ia.forth.gr
Daniel Morcuende, \inst{\ref{InstAstro_Granada}} %\orcid{} %dmorcuende@iaa.es
Alfredo Sota, \inst{\ref{InstAstro_Granada}} \orcid{0000-0002-9404-6952} %sota@iaa.es
Rumen Bachev, \inst{\ref{Astro_Sofia}} %bachevr@astro.bas.bg
Tatiana S. Grishina, \inst{\ref{StPetersburg}} \orcid{0000-0002-3953-6676} %t.s.grishina@spbu.ru
Evgenia N. Kopatskaya, \inst{\ref{StPetersburg}} \orcid{0000-0001-9518-337X} %enik1346@rambler.ru
Elena G. Larionova, \inst{\ref{StPetersburg}} \orcid{0000-0002-2471-6500} %sung2v@mail.ru
Daria A. Morozova, \inst{\ref{StPetersburg}} \orcid{0000-0002-9407-7804} %d.morozova@spbu.r
Sergey S. Savchenko, \inst{\ref{StPetersburg},\ref{Pulkovo}} \orcid{0000-0003-4147-3851} %s.s.savchenko@spbu.ru
Ekaterina V. Shishkina\inst{\ref{StPetersburg}} \orcid{0009-0002-2440-2947} %e.v.shishkina99@yandex.ru
Ivan S. Troitskiy, \inst{\ref{StPetersburg}} \orcid{0000-0002-4218-0148} %i.troitsky@spbu.ru
Yulia V. Troitskaya, \inst{\ref{StPetersburg}} \orcid{0000-0002-9907-9876} %y.troitskaya@spbu.ru
Andrey A. Vasilyev, \inst{\ref{StPetersburg}} \orcid{0000-0002-8293-0214} %andrey.vasilyev@spbu.ru
}

\institute{
Institute of Astrophysics, Foundation for Research and Technology-Hellas, GR-70013 Heraklion, Greece \label{AstroCrete} 
\and
Max-Planck-Institut f\"{u}r Radioastronomie, Auf dem H\"{u}gel 69, D-53121 Bonn, Germany \label{MPI_Bonn}
\and
NASA Marshall Space Flight Center, Huntsville, AL 35812, USA \label{NASA_Alabama}
\and
University of Maryland Baltimore County Baltimore, MD 21250, USA\label{Maryland}
\and
NASA Goddard Space Flight Center Greenbelt, MD 20771, USA \label{NASA_goddard}
\and
INAF Osservatorio Astronomico di Brera, Via E. Bianchi 46, 23807 Merate (LC), Italy \label{INAF_Merate}
\and
Space Science Data Center, Agenzia Spaziale Italiana, Via del Politecnico snc, 00133 Roma, Italy \label{SSDC_Rome} 
\and
INAF Osservatorio Astronomico di Roma, Via Frascati 33, 00078 Monte Porzio Catone (RM), Italy \label{INAF_Rome_Obs} 
\and
Instituto de Astrof\'{i}sica de Andaluc\'{i}a, IAA-CSIC, Glorieta de la Astronom\'{i}a s/n, 18008 Granada, Spain \label{InstAstro_Granada}
\and
Istituto Nazionale di Fisica Nucleare, Sezione di Padova, 35131 Padova, Italy\label{INFN-Padova}
\and
Department of Physics, University of Crete, GR-70013 Heraklion, Greece \label{AstroCrete2} 
\and
Centre for Space Research North-West University Potchefstroom, 2531, South Africa \label{Northwest}
\and
Science and Technology Institute, Universities Space Research Association, Huntsville, AL 35805, USA \label{USRA_Alabama}
\and
Institute for Astrophysical Research, Boston University, 725 Commonwealth Avenue, Boston, MA 02215, USA \label{BostonUni}
\and
St. Petersburg State University, 7/9, Universitetskaya nab., 199034 St. Petersburg, Russia \label{StPetersburg}
\and
Physics Department and McDonnell Center for the Space Sciences, Washington University in St. Louis, St. Louis, MO 63130, USA \label{StLouis}
\and
Department of Physics and Kavli Institute for Particle Astrophysics and Cosmology, Stanford University, Stanford, California 94305, USA \label{Stanford}
\and
Department of Physics and Astronomy, University of Turku, FI-20014, Finland \label{UTU}
\and
Finnish Centre for Astronomy with ESO (FINCA), Quantum, Vesilinnantie 5, FI-20014 University of Turku, Finland \label{FINCA}
\and
Aalto University Mets\"ahovi Radio Observatory, Mets\"ahovintie 114, FI-02540 Kylm\"al\"a, Finland \label{MRO}
\and
Astrophysics Research Institute, Liverpool John Moores University, Liverpool Science Park IC2, 146 Brownlow Hill, UK \label{LJMU}
\and
Institut de Radioastronomie Millim\'{e}trique, Avenida Divina Pastora, 7, Local 20, E–18012 Granada, Spain \label{IRAM}
\and
Center for Astrophysics | Harvard \& Smithsonian , 60 Garden Street, Cambridge, MA 02138 USA \label{Harvard_Smithsonian}
\and
Korea Astronomy and Space Science Institute, 776 Daedeok-daero, Yuseong-gu, Daejeon 34055, Korea \label{Korea_ASSI}
\and
University of Science and Technology, Korea, 217 Gajeong-ro, Yuseong-gu, Daejeon 34113, Korea \label{Korea_UniSciTech}
\and
Section of Astrophysics, Astronomy \& Mechanics, Department of Physics, National and Kapodistrian University of Athens, Panepistimiopolis Zografos 15784, Greece \label{Greece_UniAthens}
\and
Institute of Astronomy and NAO, Bulgarian Academy of Sciences, 1784 Sofia, Bulgaria \label{Astro_Sofia}
\and
Pulkovo Observatory, St.Petersburg, 196140, Russia \label{Pulkovo}
}

%\date{Received September 15, 1996; accepted March 16, 1997}

% \abstract{}{}{}{}{} 
% 5 {} token are mandatPeriodicity in astrophysical jets can be attributed to a plethora of mechanism such as {\color{red}add stuff}. While there are a few notable examples in X-ray binaries with jets a.k.a microquasars, identifying periodicity in light curves of active galactic nuclei (AGN) has been a much more cumbersome endeavor. 

\abstract{The origin of the high-energy emission in astrophysical jets from black holes is a highly debated issue. This is particularly true for jets from supermassive black holes, which are among the most powerful particle accelerators in the Universe. So far, the addition of new observations and new messengers have only managed to create more questions than answers. However, the newly available X-ray polarization observations promise to finally distinguish between emission models. We use extensive multiwavelength and polarization campaigns as well as state-of-the-art polarized spectral energy distribution models to attack this problem by focusing on two X-ray polarization observations of blazar BL Lacertae in flaring and quiescent $\gamma$-ray states.  We find that, regardless of the jet composition and underlying emission model, inverse-Compton scattering from relativistic electrons dominates at X-ray energies.}


\titlerunning{High-energy emission mechanism through multiwavelength polarization}
\authorrunning{Liodakis et al.}

\maketitle
%
%-------------------------------------------------------------------
\section{Introduction} \label{sec:intro}

Blazars are the brightest and most variable subclass of active galactic nuclei (AGN) with emission that spans the entire electromagnetic and, potentially, the particle spectrum \citep[e.g.,][]{Hovatta2019,Blandford2019}. This is attributed to the orientation of their jets, which are aligned with our line of sight, resulting in relativistically boosted emission and timescale compression \cite[e.g.,][]{Liodakis2018-II}. Blazars have a spectral energy distribution (SED) made up of two broad emission components. The low-energy component from radio to X-rays is produced by relativistic electrons spiraling in the magnetic field of the jet. The particle acceleration mechanism is still debated, but recent results from the Imaging X-ray Polarimetry Explorer \cite[IXPE; ][]{Weisskopf2022} favor shock acceleration \cite[e.g.,][]{Liodakis2022,Kouch2024}. The location of the synchrotron peak is often used to classify blazars into low ($\rm  \nu_{syn}<10^{14}~Hz$), intermediate ($\rm 10^{14}~Hz< \nu_{syn}<10^{15}~Hz$), and high ($\rm \nu_{syn}>10^{15}~Hz$) synchrotron peaked sources \citep{Ajello2020}. In contrast, the origin of the high-energy component (X-rays to TeV $\gamma$-rays) is uncertain. If the jet emission is dominated by relativistic electrons, the high-energy production mechanism is expected to be inverse-Compton scattering (also known as leptonic processes). If the scattered photon field consists of external photons from the accretion disk or surrounding structures, we refer to the process as external Compton (EC). If instead the photons are synchrotron from the jet's electrons, then we refer to the process as synchrotron self-Compton (SSC). On the other hand, if the emission is dominated by protons, the production mechanism for X-rays and $\gamma$-rays would likely be either proton-synchrotron or proton-proton and proton-photon interactions (also known as hadronic processes, (e.g., \citealp{Mannheim1993}). Many previous spectral models consider leptonic processes, with only very few codes capable of modeling hadronic emission. Nevertheless, both leptonic and hadronic models can fit the typical blazar spectra comparably well \citep{Boettcher2013}.

Most of the previous studies that aimed to understand the origin of the high energy component of the blazar SED favor leptonic processes. However, the presence of orphan flares \citep{Liodakis2019,deJaeger2023} and the potential association of TXS~0506+056 with a high-energy neutrino \citep{IceCube2018} leave room for potential hadronic emission. Polarization and, specifically, the novel X-ray polarization observations made possible by IXPE can give a fresh perspective on the problem. Hadronic X-ray and $\gamma$-ray emission  is typically expected to produce comparable polarization degree to the radio and optical bands, but the polarization is less variable than in the optical band \citep{Zhang2019,Zhang2024}. Leptonic processes are expected to produce high-energy radiation with either no polarization or a much lower degree of polarization than radio and optical \citep{Poutanen1994,Peirson2019,Liodakis2019-II,Peirson2022}. So far all the X-ray polarization observations have only yielded upper limits, and they were often not constraining or the optical polarization degree was low, preventing us from drawing strong conclusions \citep{Middei2023,Marshall2023,Kouch2025}. 

BL Lacertae (BL Lac), the original BL Lac object, is located at $\rm RA=22h~02m~43.2s$, $\rm Dec=+42\rm^o~16^\prime~39.9^{\prime\prime}$ and z=0.069. It is the only low-/intermediate-peaked blazar that has been observed multiple times by IXPE. Here, we attempt to combine SED modeling with the discriminatory power of multiwavelength (MWL) polarization by  focusing on the last two observations of BL Lac, which provide the best constraints on the high-energy emission processes. In Sect. \ref{sec:data} we present the multiwavelength flux and polarization observations, and in Sect. \ref{sec:sed_mod} we present the SED models we use to fit the data. In Sect. \ref{sec:discussion} we discuss our results.

\section{Multiwavelength polarization observations} \label{sec:data}

We focus on the third and fourth observations (hereafter, OBS3 and OBS4) of BL Lac by IXPE \citep{Peirson2023,Agudo2025}. For OBS3 BL Lac was in a flaring $\gamma$-ray state. \cite{Peirson2023} split the IXPE exposure in three equal parts (hereafter SEG1, SEG2, and SEG3) and found that BL Lac was in an intermediate-peak state for SEG1 and in a low peak state for SEG2 and SEG3.  Effectively, in SEG1 the 2-8~keV band has contribution from both the electron synchrotron and high-energy hump, whereas in SEG2 and SEG3 the 2-8~keV band is entirely in the high-energy hump of the SED. In SEG1 polarization was detected from the electron synchrotron component in the 2-4 keV band with a degree of about 20\%. The 2-8~keV polarization was undetected with a 99\% upper limit of 28\%. Similarly, the X-ray polarization in SEG2, SEG3, and the entire OBS3 yielded only upper limits of 23\%,  22\%, and 14.3\%, respectively. OBS4 was taken during a quiescent state in $\gamma$-rays, but an elevated state in radio, optical, and X-rays. The millimeter radio flare observed during the IXPE observation (OBS4) was the brightest flare in BL Lac over the past $\sim$40 years \citep{Agudo2025,Mondal2025}. The X-ray polarization was undetected, yielding the most stringent  3$\sigma$ upper limit so far of 7.4\%.  At the same time, the optical polarization degree reached a historical maximum of 47.6\%, making it the most polarized blazar ever observed.

For the total intensity and polarized SEDs, we used only observations taken within the span of the IXPE observation to ensure simultaneity. The datasets include observations from a large number of ground and space-based telescopes in radio, optical, X-rays, and $\gamma$-rays. These are, namely, the Belogradchik Observatory \citep{Bachev2023}, Calar alto (CAFOS), Effelsberg (QUIVER, \citealp{Krauss2003,Myserlis2018} , the {\it Fermi} gamma-ray space telescope, IRAM-30m (POLAMI, \citealp{Agudo2018}), IXPE \citep{Weisskopf2022}, KVN \citep{Kang2015} , Liverpool Telescope (MOPTOP, \citealp{Shrestha2020}), the Nordic Optical Telescope (ALFOC, \citealp{{Nilsson2018}}), NuSTAR, Sierra Nevada Observatory (DIPOL-1, \citealp{{otero2024}}), Perkins (PRISM), Skinakas observatory (RoboPol, \citealp{Panopoulou2015,Ramaprakash2019,Blinov2021}), SMA (SMAPOL, Myserlis et al., in prep.),  St. Petersburg (LX-200), the Niel Gehrels {\it Swift} observatory, and XMM-Newton. The observations and data reduction procedures are discussed in detail in (\citealp{Peirson2023,Agudo2025}). In order to directly compare the X-ray polarization observations, we used the averaged radio and optical polarization degree over the duration of the IXPE observation.

For the $\gamma$-ray observations, we used the standard analysis with {\it Fermitools} v1.2.23 and Pass8 P8R3 source events. We collected all the \textit{Fermi}-LAT data of BL Lac in the energy range between $>100$~MeV and 300~GeV from the LAT data server\footnote{\url{https://fermi.gsfc.nasa.gov/ssc/data/access/lat/}} and within a region of interest (ROI) of 15 deg. We filtered all the data with a zenith angle $>90^{\circ}$ to avoid contamination from the Earth's limb. We also employed the recommended Galactic and diffuse emission components\footnote{\url{https://fermi.gsfc.nasa.gov/ssc/data/access/lat/BackgroundModels.html}}, i.e., \texttt{gll\_iem\_v07} and \texttt{iso\_P8R3\_SOURCE\_V3\_v1}, respectively.

To account for the emission of all nearby sources, we included in our model all the sources within the 15-degree radius ROI, plus those within an additional annular region of radius 10 deg. The spectral parameters of all variable sources within the ROI with TS > 25 (as defined in the 4FGL catalog, \citealp{Abdollahi2020}) were left as free parameters, while those contained in the annular region or not fulfilling these conditions were fixed to the catalog values. Weak sources with TS < 4 were removed from the model. The normalization of the diffuse components was allowed to vary. Lastly, we performed a binned likelihood analysis of BL Lac’s data with the final model, computing the spectrum and SED, modeling each flux point with a power-law shape. The binned analysis is recommended for sources close to bright background sources such as BL Lac. We verified that an unbinned analysis yields consistent results within uncertainties.

\section{SED models}\label{sec:sed_mod}

\begin{figure}
\centering
 \resizebox{\hsize}{!}{\includegraphics[width=\textwidth]{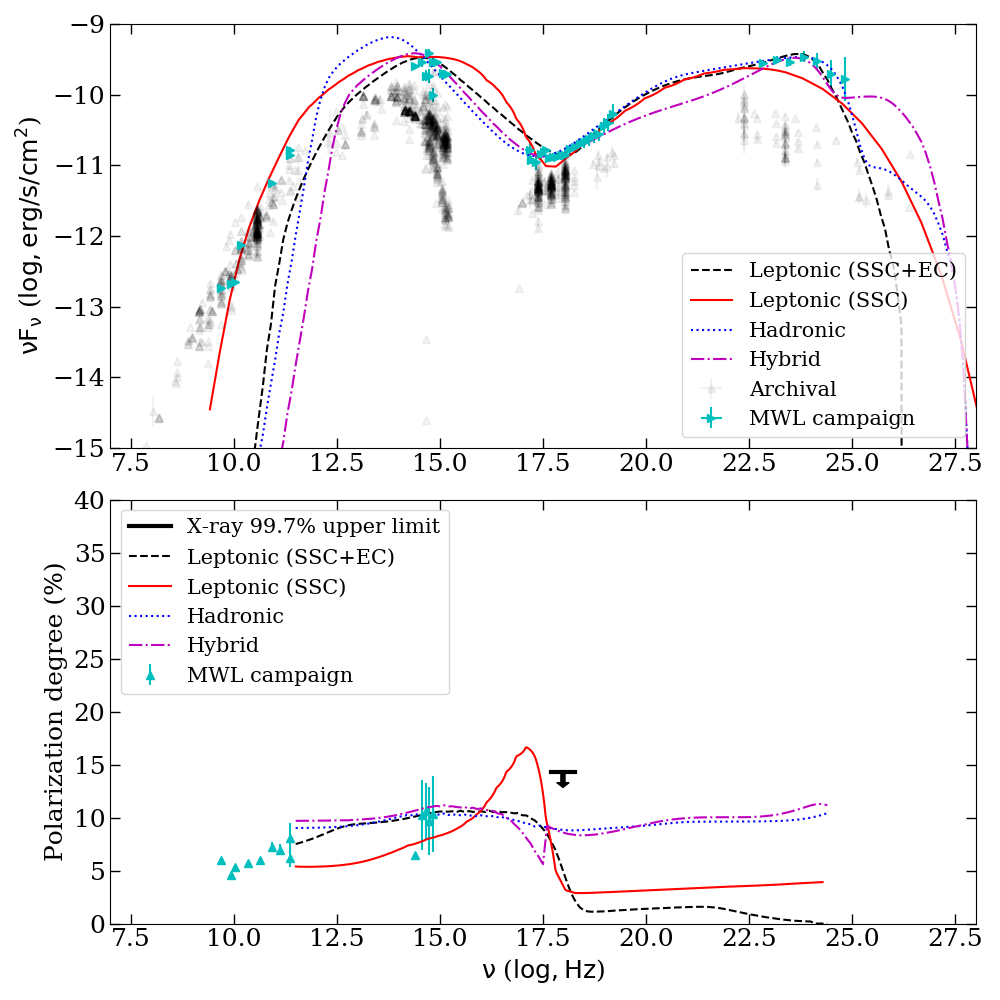}}
 \caption{Total intensity (top panel) and polarization (bottom panel) SED modeling for OBS3. The simultaneous observations from the MWL campaign are shown in cyan. The archival observations from the Space Science Data Center are shown in black. For both flux and polarization degree, we show the median and standard deviation in different frequencies within the IXPE observation. The lines corresponding to the different emission models are listed in the legend.}
    \label{plt:SED_OBS3}
\end{figure}

\begin{figure}
\centering
 \resizebox{\hsize}{!}{\includegraphics[width=\textwidth]{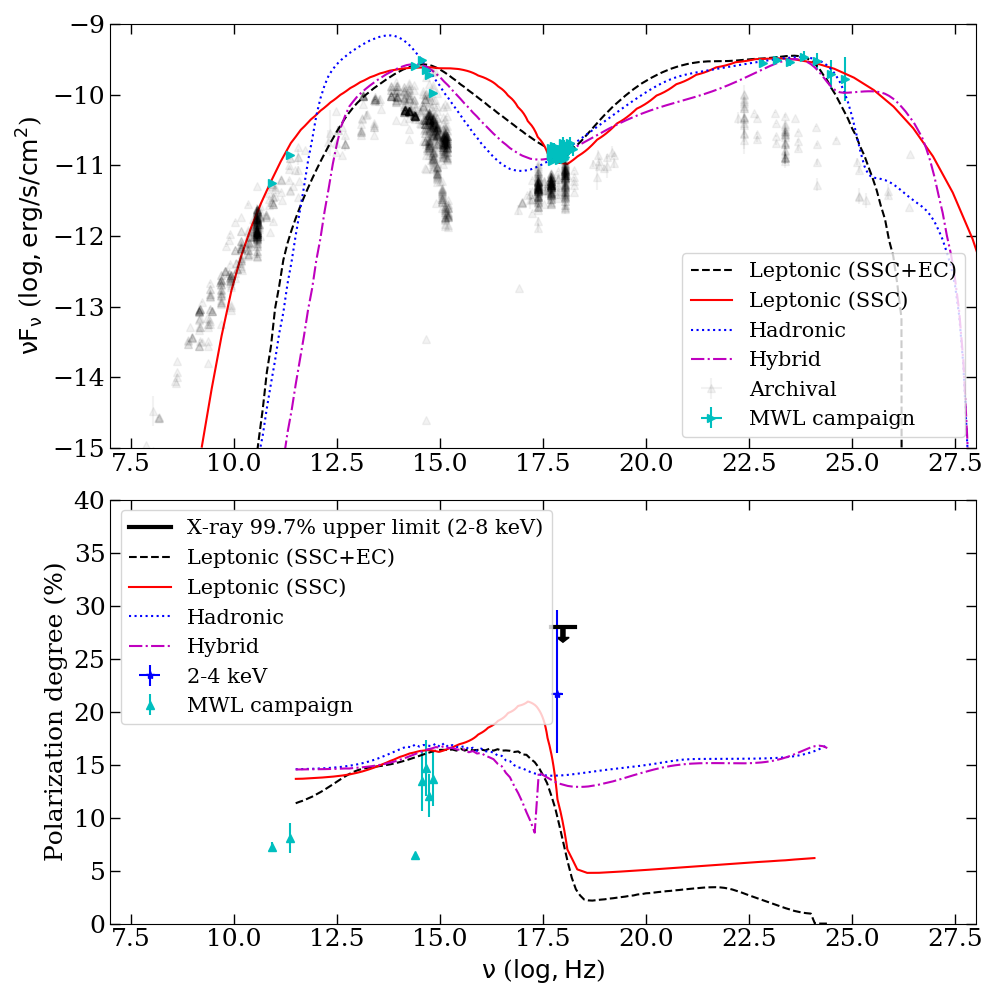}}
 \caption{Same as Fig. \ref{plt:SED_OBS3}, but for SEG1. }
    \label{plt:SED_SEG1}
\end{figure}

\begin{figure}
\centering
 \resizebox{\hsize}{!}{\includegraphics[width=\textwidth]{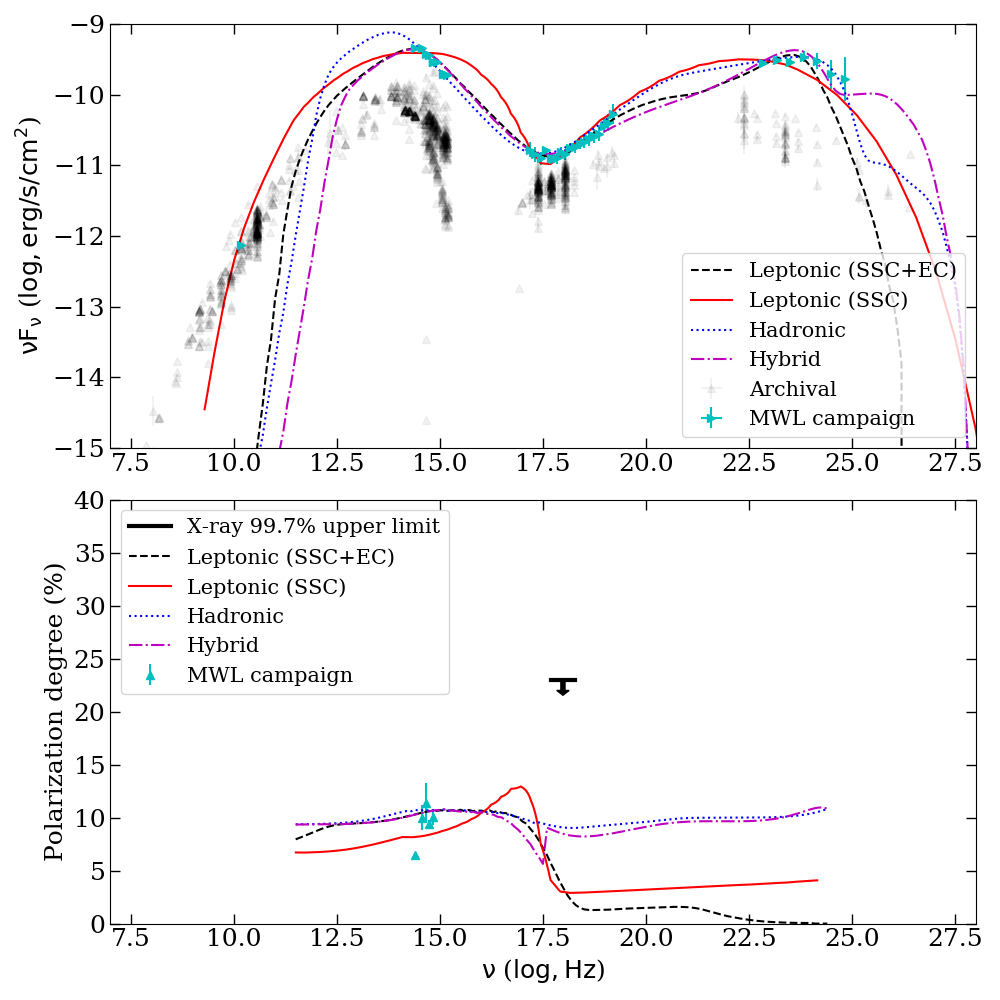}}
 \caption{Same as Fig. \ref{plt:SED_OBS3}, but for SEG2. }
    \label{plt:SED_SEG2}
\end{figure}

\begin{figure}
\centering
 \resizebox{\hsize}{!}{\includegraphics[width=\textwidth]{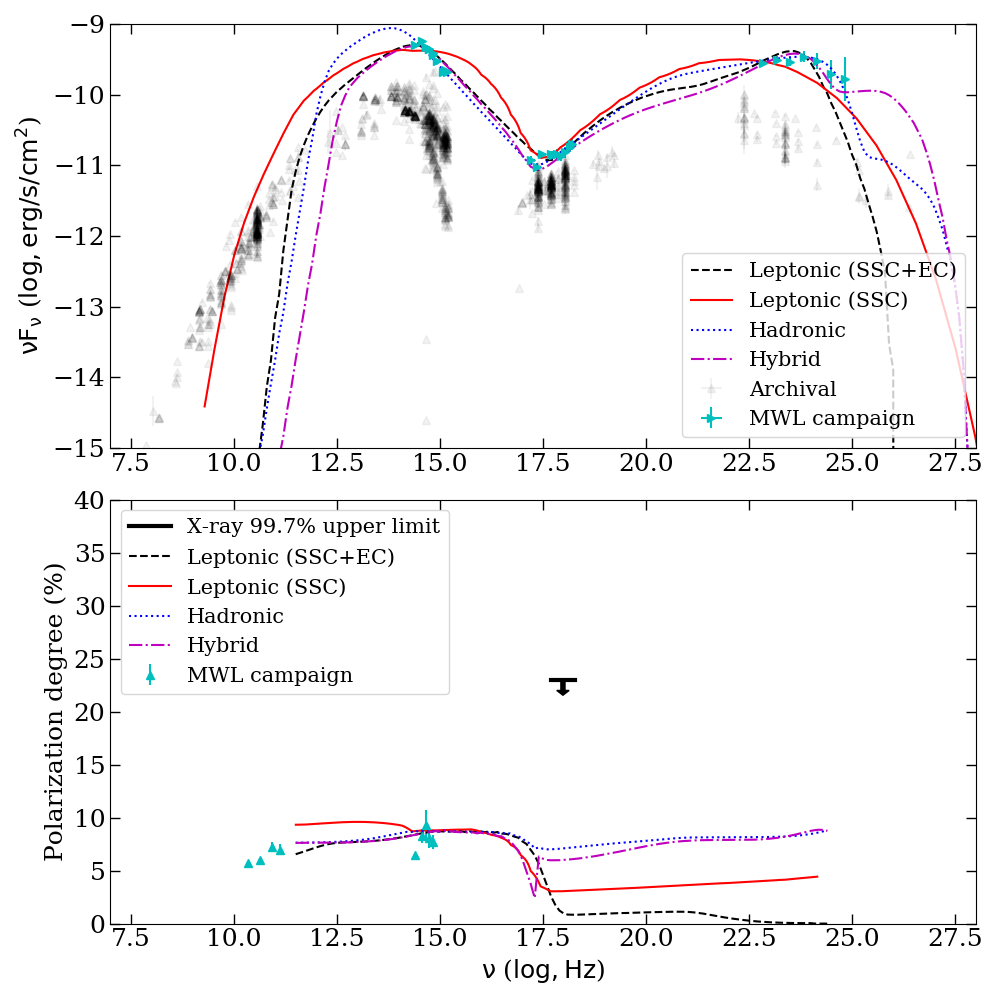}}
 \caption{Same as Fig. \ref{plt:SED_OBS3}, but for SEG3. }
    \label{plt:SED_SEG3}
\end{figure}

\begin{figure}
\centering
 \resizebox{\hsize}{!}{\includegraphics[width=\textwidth]{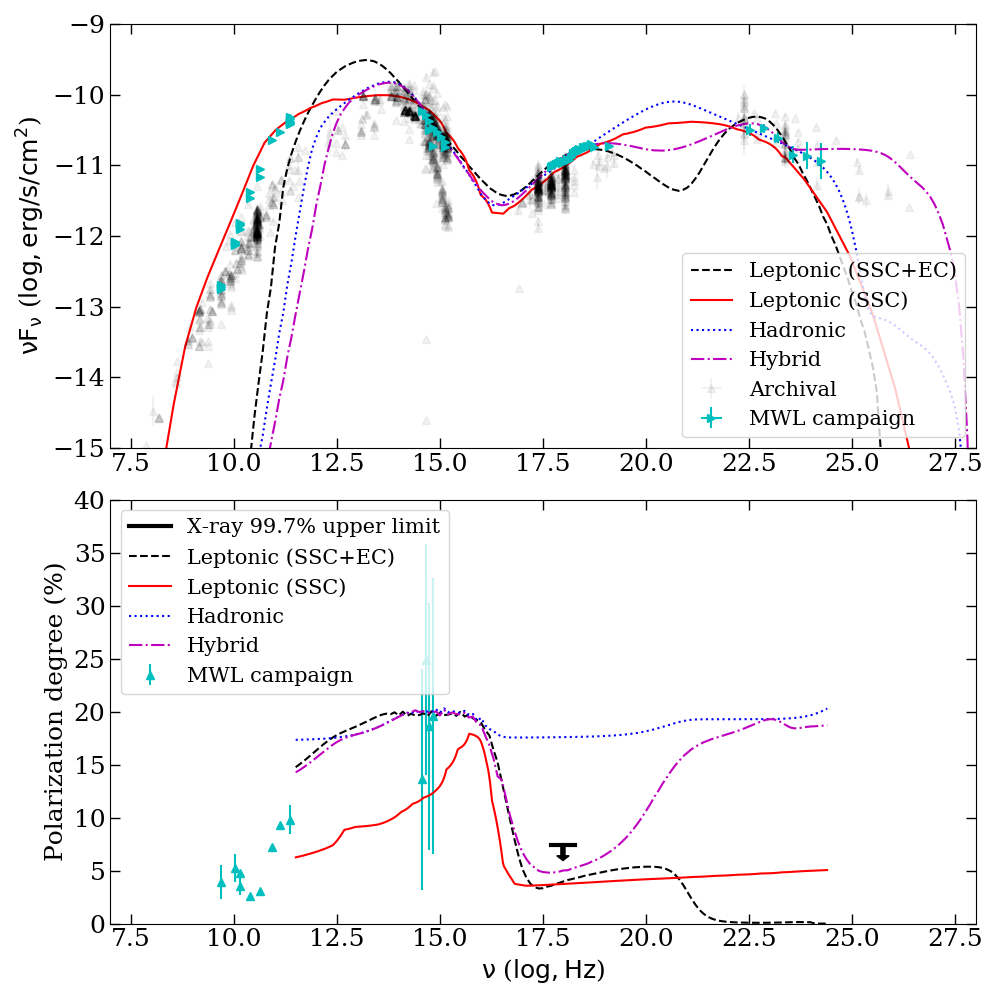}}
 \caption{Total intensity (top panel) and polarization (bottom panel) SED modeling for OBS4. Symbols and colors are the same as in Fig. \ref{plt:SED_OBS3}.}
    \label{plt:SED_OBS4}
\end{figure}

We consider four models for the high-energy spectral component: a pure SSC model, a SSC + EC leptonic model, a hybrid model with SSC and hadronic processes, and a pure hadronic model. The low-energy spectral component consists of electron synchrotron in all models. The pure SSC model uses the code developed by \cite{Peirson2019}. This multi-zone code considers inhomogeneous magnetic field distribution and calculates both electron synchrotron and SSC flux and polarization. It thus naturally includes multi-zone effects such as energy stratification and depolarization. The synchrotron self-absorption process is not included. The other three models use the one-zone spectral fitting code, developed by \cite{Boettcher2013} and post-processed by the polarization code developed by \cite{Zhang2013}, and includes the semi-analytical depolarization in \cite{Zhang2024}. The spectral fitting code considers synchrotron self absorption, synchrotron for electrons and protons, SSC and EC, pair synchrotron from hadronic cascades, as well as all relevant radiative cooling. The external photon field is a blackbody spectrum at temperature $T$ with a density $u_{ext}$ that is uniform in the emission blob. The polarization post-processing adds multi-zone depolarization effects. We find that the fitting result is reasonable without the energy stratification, so it is not included in the polarization post-processing. For all the polarization models, the synchrotron component does not include synchrotron self absorption, and the SSC components are in the Thomson regime (i.e., no Klein-Nishina effects are included). Therefore, any polarization estimations below $\sim10^{12}$~Hz or above $\sim10^{25}$~Hz should not be considered reliable. Key fitting parameters for all four models are shown in Tables \ref{tab:ssc}, \ref{tab:ssc_ec}, \ref{tab:ssc_hadronic}, and \ref{tab:hadronic} for OBS3, SEG1, SEG2, SEG3, and OBS4. Figures \ref{plt:SED_OBS3}, \ref{plt:SED_SEG1}, \ref{plt:SED_SEG2}, \ref{plt:SED_SEG3}, and \ref{plt:SED_OBS4} show the archival and simultaneous observations, as well as the fitting models for OBS3, SEG1, SEG2, SEG3, and OBS4, respectively. The individual emission components and associated polarization degree for the different models are shown in Appendices \ref{app:SED} and \ref{app:SPD}.

\section{Discussion and conclusions}\label{sec:discussion}

We used simultaneous observations from IXPE's MWL campaigns \citep{Peirson2023,Agudo2025} to study the total intensity and polarization SED of BL Lac in different synchrotron peak and $\gamma$-ray flux states. We used the optical polarization as a benchmark for the ordering of magnetic fields and aimed to simultaneously model the MWL spectrum and polarization degree. We find that both the leptonic models and the hybrid hadronic model considered here can satisfy the MWL and polarization observations, except the X-ray constraints. For the models to satisfy the X-ray polarization upper limits, the X-ray emission in BL Lac, and consequently blazars in general, they must be dominated by inverse-Compton scattering from the relativistic electrons. Our results thus provide stringent constraints on the origin of the high-energy emission in jets.

All the one-zone model curves (SSC+EC, hybrid, and hadronic) are below the radio to far-infrared data points. This is because the radio to far-infrared emission can come from a much larger portion of the blazar jet rather than merely the blazar zone. On the other hand, the multi-zone pure SSC model can consider both the blazar zone and the large-scale jet. Hence, it fits those data points reasonably well. However, the pure SSC model generally misses the $\gamma$-ray data, suggesting the need for an EC component. 

We note that the hybrid model yields a very hard spectrum in the TeV bands, due to the strong contribution from hadronic cascades. Current Cherenkov telescopes can test this scenario. If this component is not detected, then the hybrid model will be strongly disfavored. This also highlights the potential synergies between IXPE and future Cherenkov telescopes (e.g., CTAO; \citealp{Boisson2021}).

All the models can satisfy the X-ray polarization upper limits for OBS3, SEG1, SEG2, and SEG3. Nonetheless, the hybrid and hadronic models are disfavored. Based on our modeling for SEG1, the hadronic and hybrid models predict a roughly constant polarization degree across the IXPE band. In that case, one would expect the X-ray polarization analysis for the 2-4~keV and 2-8~keV bands to yield the same results. Instead, we observe a detection in the 2-4~keV band and a non-detection in the 2-8~keV band. This would imply that the upper half of the IXPE band (4-8~keV) is dominated by a different process than the lower half. Additionally, the 2-4~keV polarization is not detected in SEG2 and SEG3, showing that the polarization is variable within a few days in the observer's frame, corresponding to roughly one month in the comoving frame of the blazar zone. The proton energy that produces the X-ray emission via either proton synchrotron or hadronic cascades is a few PeV. Both the flux and polarization variability are controlled by magnetic field and particle evolution. If the magnetic field changes fast enough, the flux variability for electron synchrotron and proton synchrotron can be comparable. However, due to the faster cooling of electrons, changes in the magnetic field will significantly enhance and/or reduce local electron synchrotron cooling and, thus, the electron particle energy distribution, which is reflected in a comparably fast change in the polarization. On the other hand, protons cool much more slowly; therefore, changes in the magnetic field hardly alter the proton distribution. Consequently, the hadronic polarization variability should always be slower than the electron synchrotron polarization variations  \citep{Zhang2016,Zhang2019,Zhang2024}. Based on the hybrid and hadronic model parameters (see Tables \ref{tab:ssc_hadronic} and \ref{tab:hadronic}), the cooling timescale for protons is $\gtrsim 30$~yr in the comoving frame. If the X-ray and $\gamma$-ray emission are of hadronic origin, then the hadronic polarization is unlikely to change within OBS3 as shown in \citet{Zhang2016}. Therefore, hybrid and hadronic models cannot adequately and consistently explain OBS3, SEG1, SEG2, and SEG3. We note that none of the models can fully explain the high 2-4~keV polarization. We suggest that it may result from a highly inhomogeneous synchrotron contribution from a localized region with very ordered magnetic fields (e.g., a shock), which is not fully captured by the multi-zone pure SSC model or the semi-analytical multi-zone polarization post-processing of the SSC+EC model. The derived parameters for both leptonic models roughly yield an equipartition between the electron energy and magnetic energy densities (see Tables \ref{tab:ssc} and \ref{tab:ssc_ec}). Since the energy in protons is unconstrained in the leptonic models, it is unclear how magnetized the blazar zone is or what drives the particle acceleration. But given the high 2-4~keV polarization in SEG1, the emission region is likely energy stratified.

Our strongest constraints come from OBS4, which combines the lowest X-ray upper limit and the highest optical polarization degree. We find that a pure hadronic model produces an X-ray polarization degree that is too high compared to the observations and, therefore, can be ruled out. Instead, the X-ray polarization upper limit can be satisfied with models invoking SSC+EC or SSC+hadronic emission, where SSC dominates the IXPE band. Whether the radiative output at higher energies is dominated by leptonic or hadronic processes is still a mystery. Interestingly, in our modeling, leptonic processes continue to have a low polarization degree at higher ($\rm >MeV$) energies, whereas hadronic processes are typically a factor of two higher. Simultaneous X-ray (IXPE) and $\gamma$-ray polarization observations with future missions (e.g., COSI, \citealp{Tomsick2023}; AMEGO, \citealp{Rani2019}) could allow us to differentiate between emission scenarios.

\begin{acknowledgements}
We thank the anonymous referee for helpful comments that helped improve the paper. The Imaging X-ray Polarimetry Explorer (IXPE) is a joint US and Italian mission.  The US contribution is supported by the National Aeronautics and Space Administration (NASA) and led and managed by its Marshall Space Flight Center (MSFC), with industry partner Ball Aerospace (contract NNM15AA18C)---now, BAE Systems.  The Italian contribution is supported by the Italian Space Agency (Agenzia Spaziale Italiana, ASI) through contract ASI-OHBI-2022-13-I.0, agreements ASI-INAF-2022-19-HH.0 and ASI-INFN-2017.13-H0, and its Space Science Data Center (SSDC) with agreements ASI-INAF-2022-14-HH.0 and ASI-INFN 2021-43-HH.0, and by the Istituto Nazionale di Astrofisica (INAF) and the Istituto Nazionale di Fisica Nucleare (INFN) in Italy. This research used data products provided by the IXPE Team (MSFC, SSDC, INAF, and INFN) and distributed with additional software tools by the High-Energy Astrophysics Science Archive Research Center (HEASARC), at NASA Goddard Space Flight Center (GSFC). Based on observations obtained with XMM-Newton, an ESA science mission with instruments and contributions directly funded by ESA Member States and NASA. We acknowledge the use of public data from the Swift data archive. Some of the data are based on observations collected at the Observatorio de Sierra Nevada; which is owned and operated by the Instituto de Astrof\'isica de Andaluc\'ia (IAA-CSIC); and at the Centro Astron\'{o}mico Hispano en Andalucía (CAHA); which is operated jointly by Junta de Andaluc\'{i}a and Consejo Superior de Investigaciones Cient\'{i}ficas (IAA-CSIC). 
The Perkins Telescope Observatory, located in Flagstaff, AZ, USA, is owned and operated by Boston University. This research was partially supported by the Bulgarian National Science Fund of the Ministry of Education and Science under grants KP-06-H68/4 (2022) and KP-06-H88/4 (2024). The Liverpool Telescope is operated on the island of La Palma by Liverpool John Moores University in the Spanish Observatorio del Roque de los Muchachos of the Instituto de Astrofisica de Canarias with financial support from the UKRI Science and Technology Facilities Council (STFC) (ST/T00147X/1). This research has made use of data from the RoboPol program, a collaboration between Caltech, the University of Crete, IA-FORTH, IUCAA, the MPIfR, and the Nicolaus Copernicus University, which was conducted at Skinakas Observatory in Crete, Greece. The data in this study include observations made with the Nordic Optical Telescope, owned in collaboration by the University of Turku and Aarhus University, and operated jointly by Aarhus University, the University of Turku and the University of Oslo, representing Denmark, Finland and Norway, the University of Iceland and Stockholm University at the Observatorio del Roque de los Muchachos, La Palma, Spain, of the Instituto de Astrofisica de Canarias. The data presented here were obtained in part with ALFOSC, which is provided by the Instituto de Astrof\'{\i}sica de Andaluc\'{\i}a (IAA) under a joint agreement with the University of Copenhagen and NOT. The Submillimeter Array (SMA) is a joint project between the Smithsonian Astrophysical Observatory and the Academia Sinica Institute of Astronomy and Astrophysics and is funded by the Smithsonian Institution and the Academia Sinica. Maunakea, the location of the SMA, is a culturally important site for the indigenous Hawaiian people; we are privileged to study the cosmos from its summit. 
The POLAMI observations reported here were carried out at the IRAM 30m Telescope. IRAM is supported by INSU/CNRS (France), MPG (Germany) and IGN (Spain).  The KVN is a facility operated by the Korea Astronomy and Space Science Institute. The KVN operations are supported by KREONET (Korea Research Environment Open NETwork) which is managed and operated by KISTI (Korea Institute of Science and Technology Information).  Partly based on observations with the 100-m telescope of the MPIfR (Max-Planck-Institut f\"ur Radioastronomie) at Effelsberg. Observations with the 100-m radio telescope at Effelsberg have received funding from the European Union’s Horizon 2020 research and innovation programme under grant agreement No 101004719 (ORP). The IAA-CSIC co-authors acknowledge financial support from the Spanish "Ministerio de Ciencia e Innovaci\'{o}n" (MCIN/AEI/ 10.13039/501100011033) through the Center of Excellence Severo Ochoa award for the Instituto de Astrof\'{i}isica de Andaluc\'{i}a-CSIC (CEX2021-001131-S), and through grants PID2019-107847RB-C44 and PID2022-139117NB-C44. I.L was supported by the NASA Postdoctoral Program at the Marshall Space Flight Center, administered by Oak Ridge Associated Universities under contract with NASA.
B. A.-G., I.L were funded by the European Union ERC-2022-STG - BOOTES - 101076343. Views and opinions expressed are however those of the author(s) only and do not necessarily reflect those of the European Union or the European Research Council Executive Agency. Neither the European Union nor the granting authority can be held responsible for them. HZ is supported by NASA under award number 80GSFC21M0002. HZ's work is supported by Fermi GI program cycle 16 under the award number 22-FERMI22-0015. This work has been partially supported by the ASI-INAF program I/004/11/4. The research at Boston University was supported in part by National Science Foundation grant AST-2108622, NASA Fermi Guest Investigator grant 80NSSC23K1507, NASA NuSTAR Guest Investigator grant 80NSSC24K0547, and NASA Swift Guest Investigator grant 80NSSC23K1145. This work was supported by NSF grant AST-2109127. E. L. was supported by Academy of Finland projects 317636 and 320045. We acknowledge funding to support our NOT observations from the Finnish Centre for Astronomy with ESO (FINCA), University of Turku, Finland (Academy of Finland grant nr 306531).
S. Kang, S.-S. Lee, W. Y. Cheong, S.-H. Kim, and H.-W. Jeong were supported by the National Research Foundation of Korea (NRF) grant funded by the Korea government (MIST) (2020R1A2C2009003, RS-2025-00562700). CC acknowledges support by the European Research Council (ERC) under the HORIZON ERC Grants 2021 programme under grant agreement No. 101040021. This work was supported by JST, the establishment of university fellowships towards the creation of science technology innovation, Grant Number JPMJFS2129.  
D.B. acknowledge support from the European Research Council (ERC) under the European Unions Horizon 2020 research and innovation program under grant agreement No.~771282. J.O.-S. acknowledges founding from the Istituto Nazionale di Fisica Nucleare (INFN) Cap. U.1.01.01.01.009.
\end{acknowledgements}

\bibliographystyle{aa} % style aa.bst
\bibliography{sample631} % your references Yourfile.bib
\newpage

\begin{appendix}

\section{Spectral energy distribution models}\label{app:SED}

\begin{figure}[H]
\centering
 \resizebox{\hsize}{!}{\includegraphics[width=\textwidth]{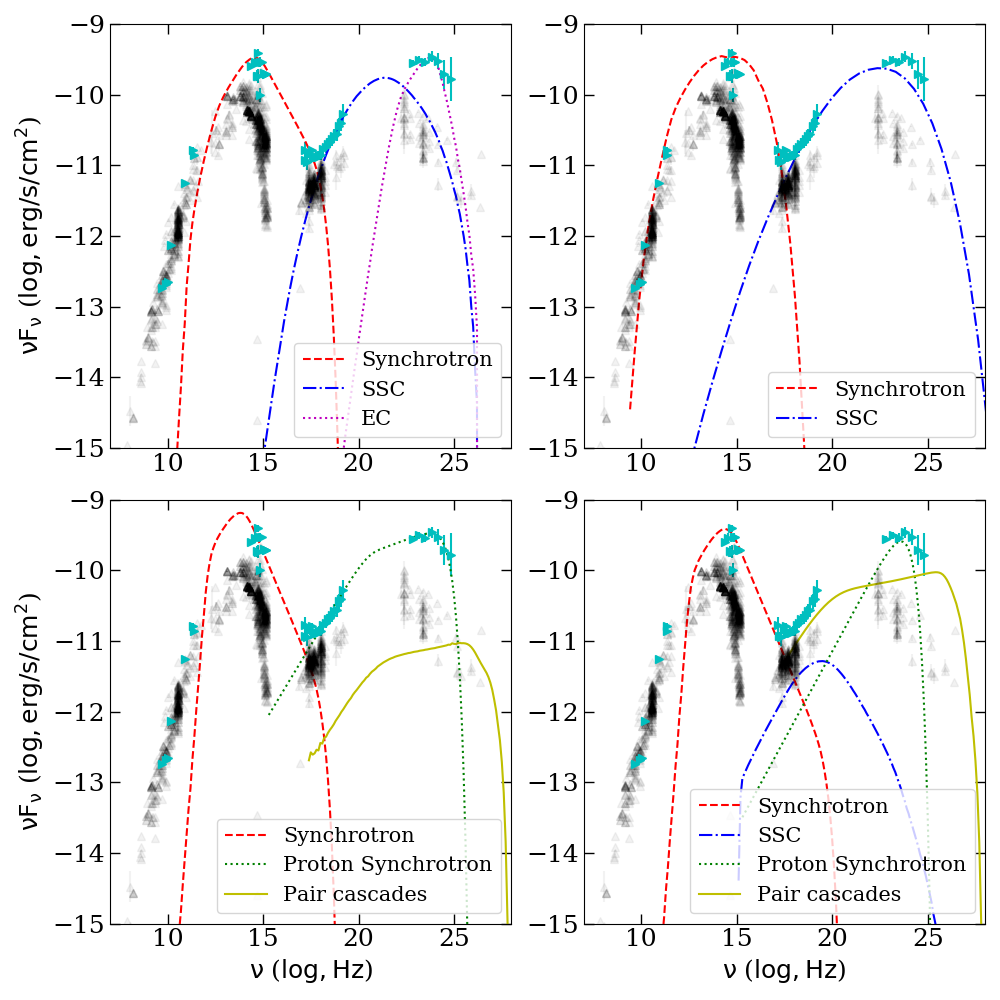}}
 \caption{Spectral energy distribution modeling for OBS3: The one-zone leptonic model (top left), the multi-zone leptonic model (top right), the hadronic model (bottom left), and the hybrid model (bottom right). The simultaneous observations from the MWL campaign are shown in cyan. The archival observations from the Space Science Data Center are shown in black.  The lines corresponding to the different emission components are listed in the legend.}
    \label{plt:SED_OBS3_indi}
\end{figure}

\begin{figure}[H]
\centering
 \resizebox{\hsize}{!}{\includegraphics[width=\textwidth]{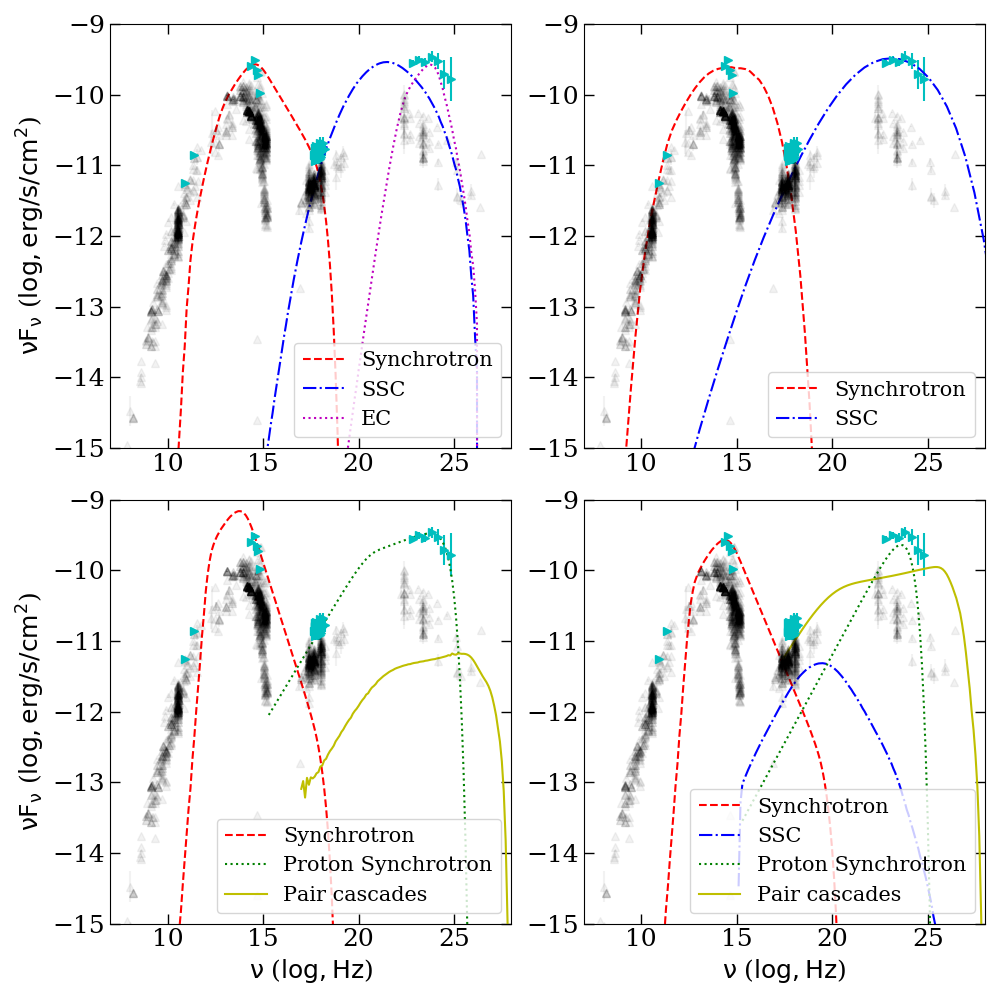}}
 \caption{Same as Fig. \ref{plt:SED_OBS3_indi}, but for SEG1.}
    \label{plt:SED_SEG1_indi}
\end{figure}

\begin{figure}[H]
\centering
 \resizebox{\hsize}{!}{\includegraphics[width=\textwidth]{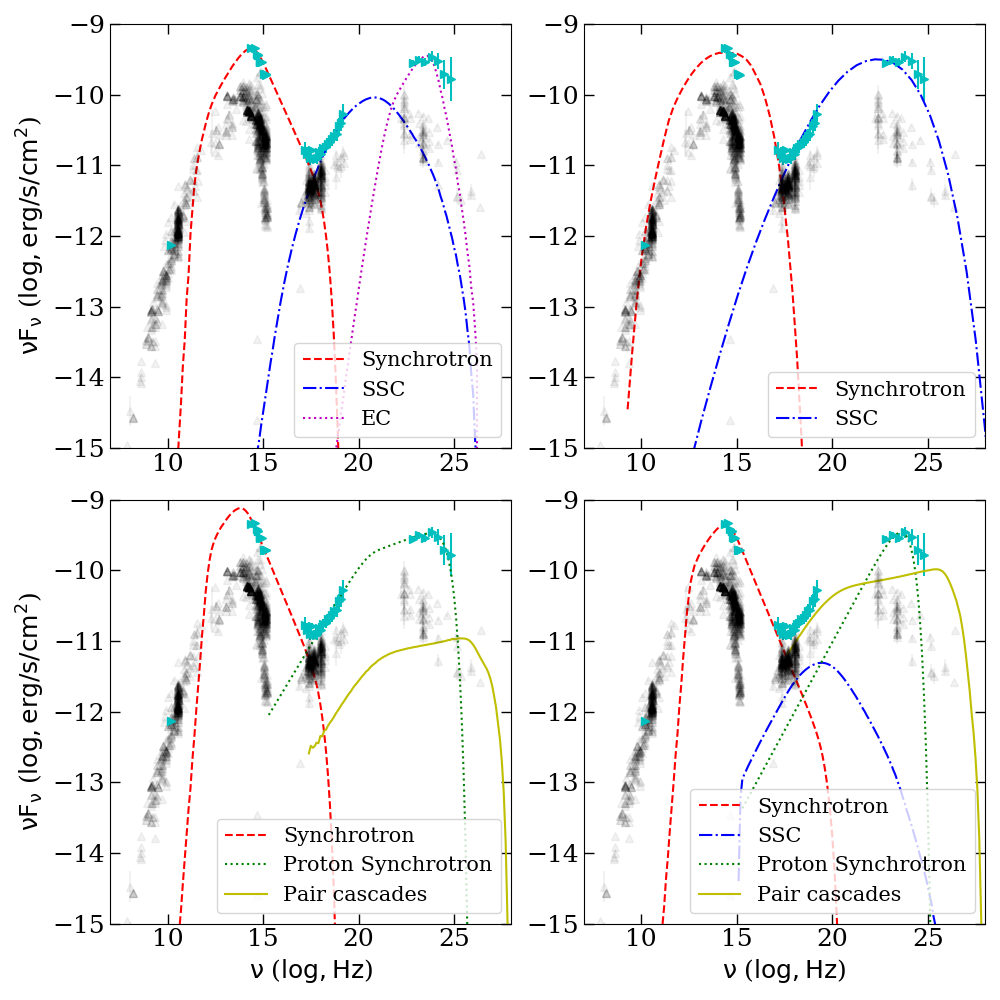}}
 \caption{Same as Fig. \ref{plt:SED_OBS3_indi}, but for SEG2. }
    \label{plt:SED_SEG2_indi}
\end{figure}

\begin{figure}[H]
\centering
 \resizebox{\hsize}{!}{\includegraphics[width=\textwidth]{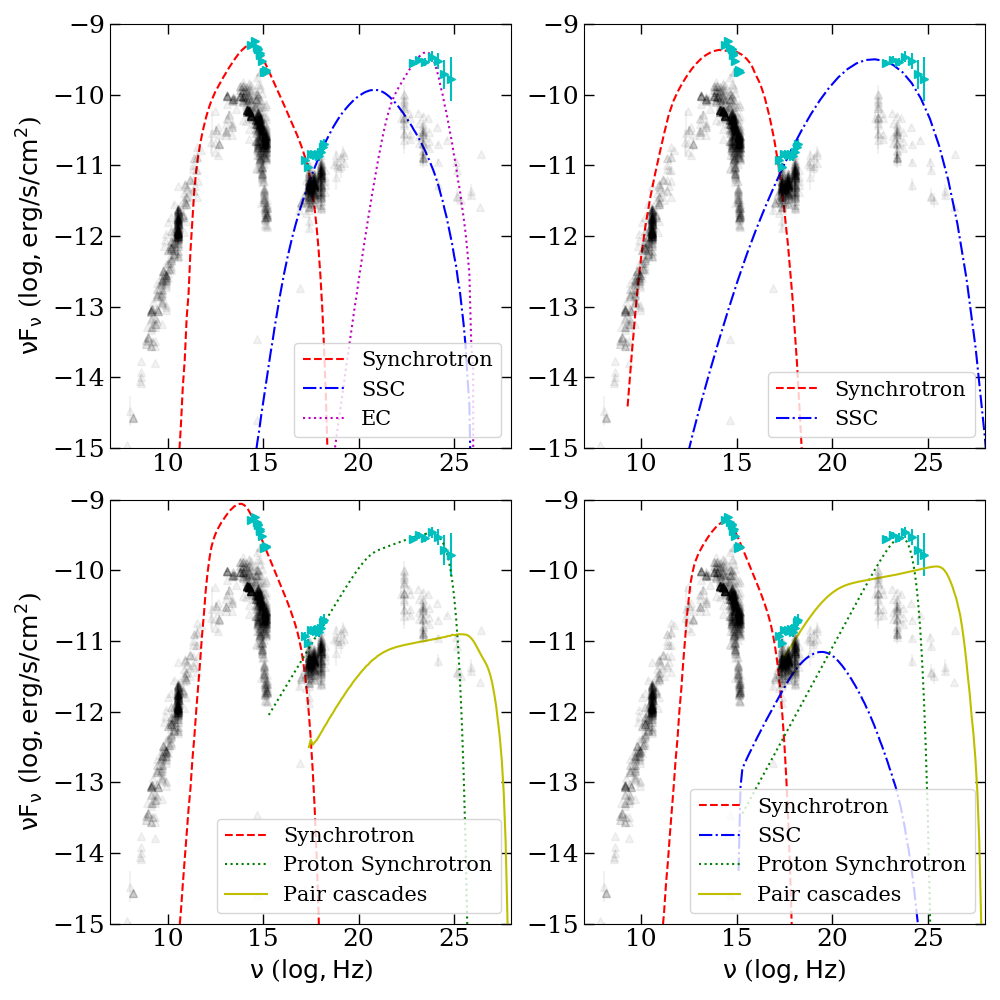}}
 \caption{Same as Fig. \ref{plt:SED_OBS3_indi}, but for SEG3. }
    \label{plt:SED_SEG3_indi}
\end{figure}

\begin{figure}[H]
\centering
 \resizebox{\hsize}{!}{\includegraphics[width=\textwidth]{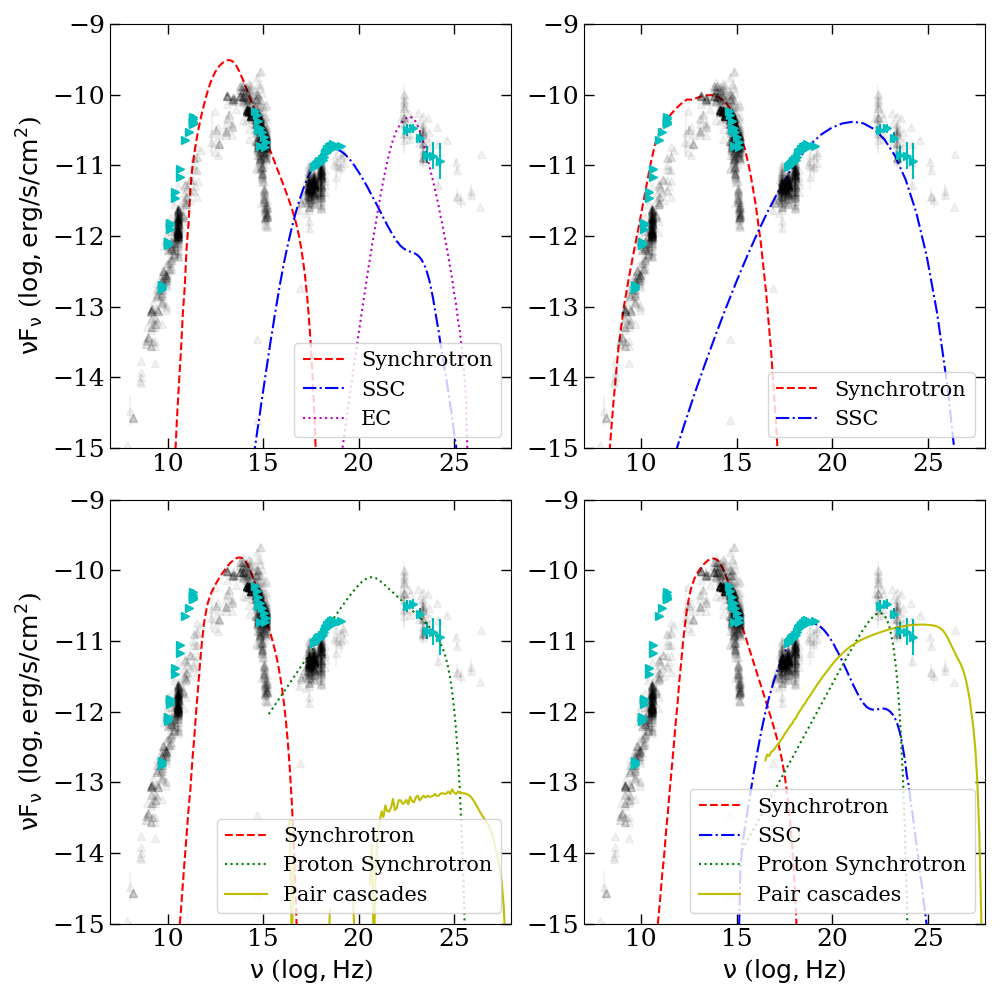}}
 \caption{Same as Fig. \ref{plt:SED_OBS3_indi} but for  OBS4.}
    \label{plt:SED_OBS4_indi}
\end{figure}

\newpage

\section{Spectral polarization models}\label{app:SPD}

\begin{figure}[H]
\centering
 \resizebox{\hsize}{!}{\includegraphics[width=\textwidth]{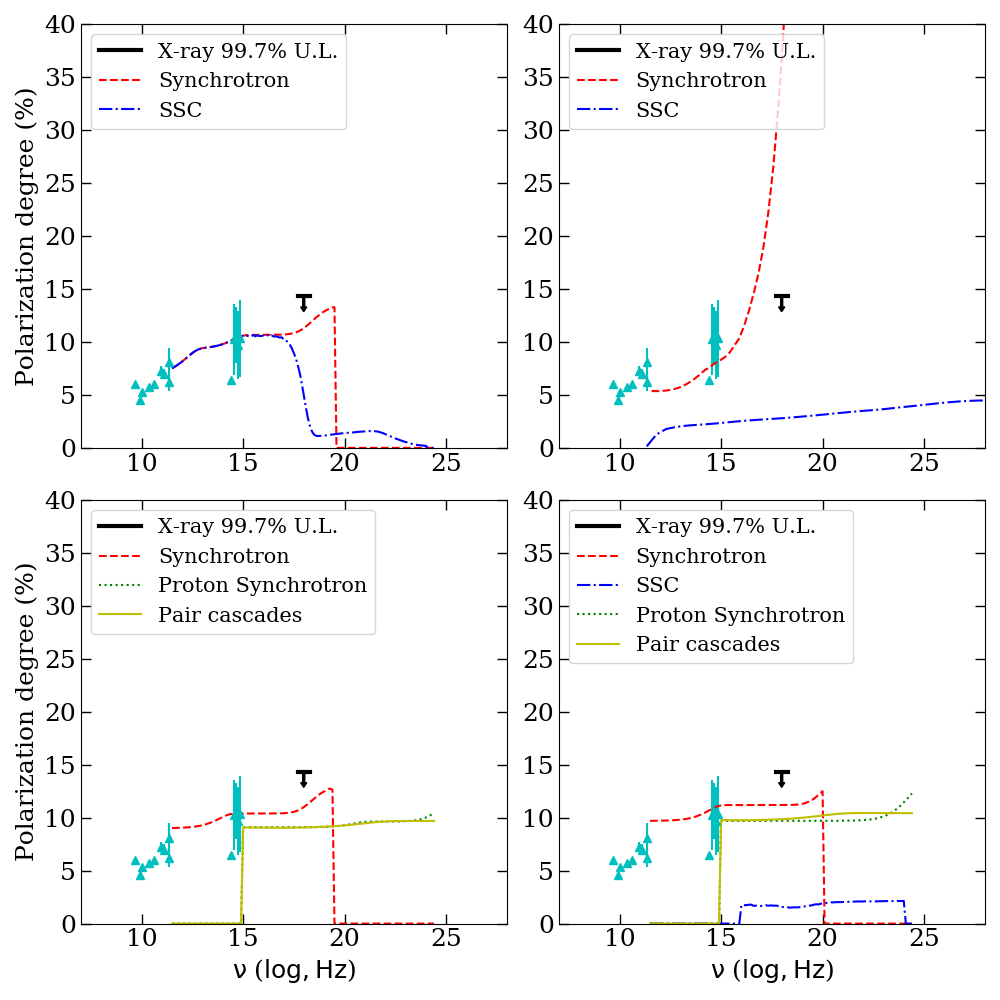}}
 \caption{Spectral polarization modeling for OBS3: The one-zone leptonic model (top left), the multi-zone leptonic model (top right), the hadronic model (bottom left), and the hybrid model (bottom right). The simultaneous observations from the MWL campaign are shown in cyan. The archival observations from the Space Science Data Center are shown in black. The lines corresponding to the different emission components are listed in the legend.}
    \label{plt:SPD_OBS3_indi}
\end{figure}

\begin{figure}[H]
\centering
 \resizebox{\hsize}{!}{\includegraphics[width=\textwidth]{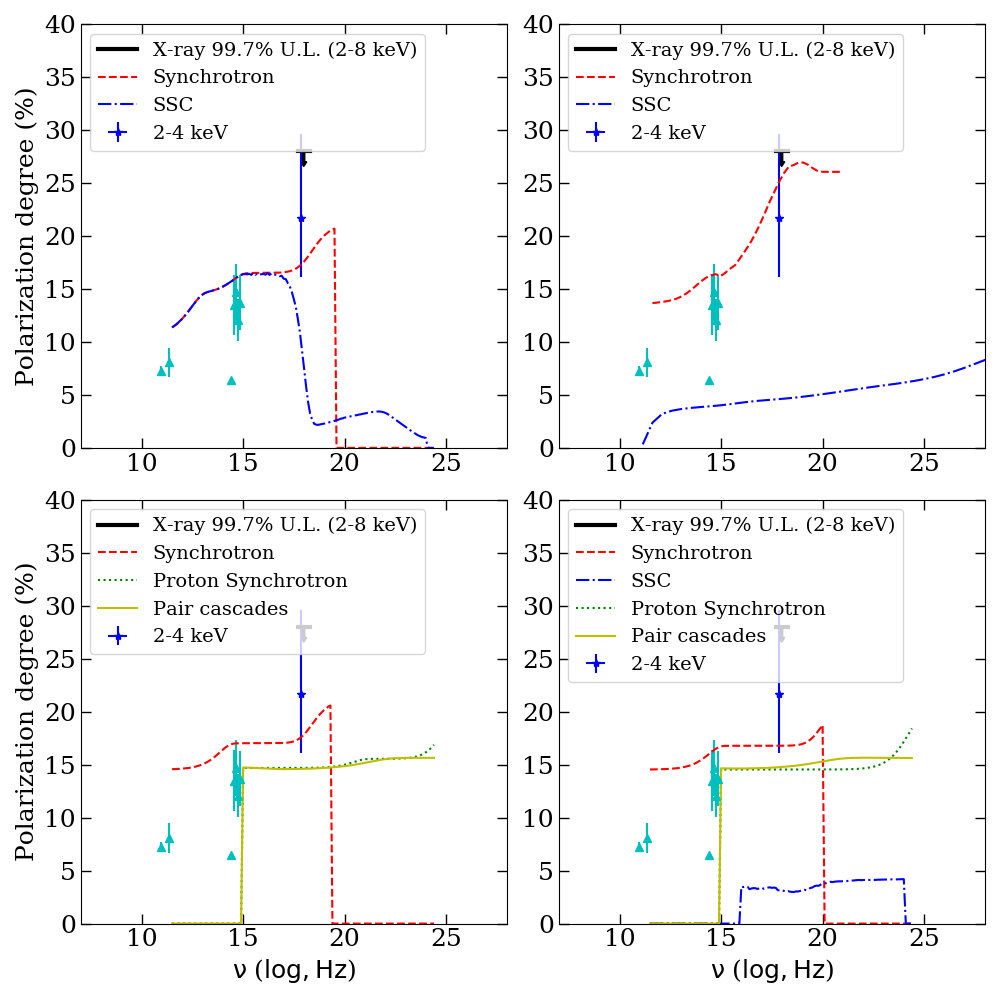}}
 \caption{Same as Fig. \ref{plt:SPD_OBS3_indi}, but for SEG1. }
    \label{plt:SPD_SEG1_indi}
\end{figure}

\begin{figure}[H]
\centering
 \resizebox{\hsize}{!}{\includegraphics[width=\textwidth]{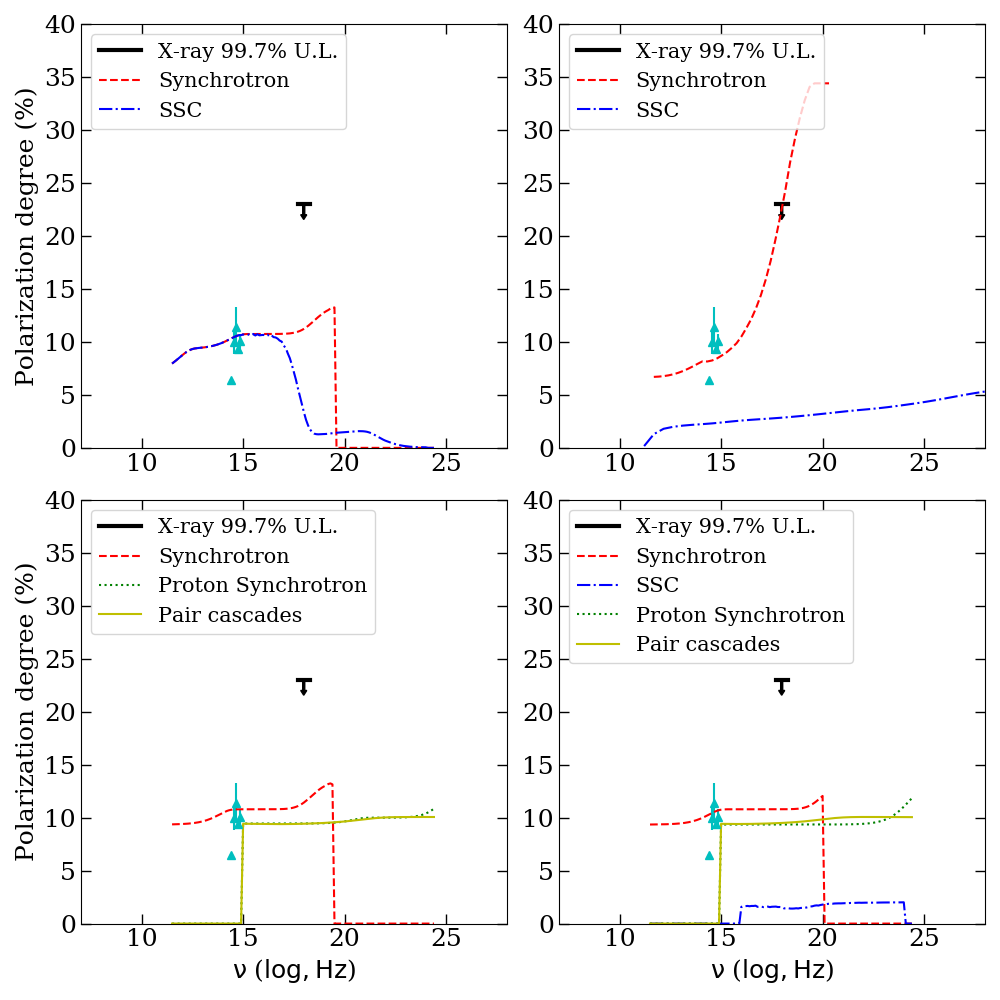}}
 \caption{Same as Fig. \ref{plt:SPD_OBS3_indi}, but for SEG2. }
    \label{plt:SPD_SEG2_indi}
\end{figure}

\begin{figure}[H]
\centering
 \resizebox{\hsize}{!}{\includegraphics[width=\textwidth]{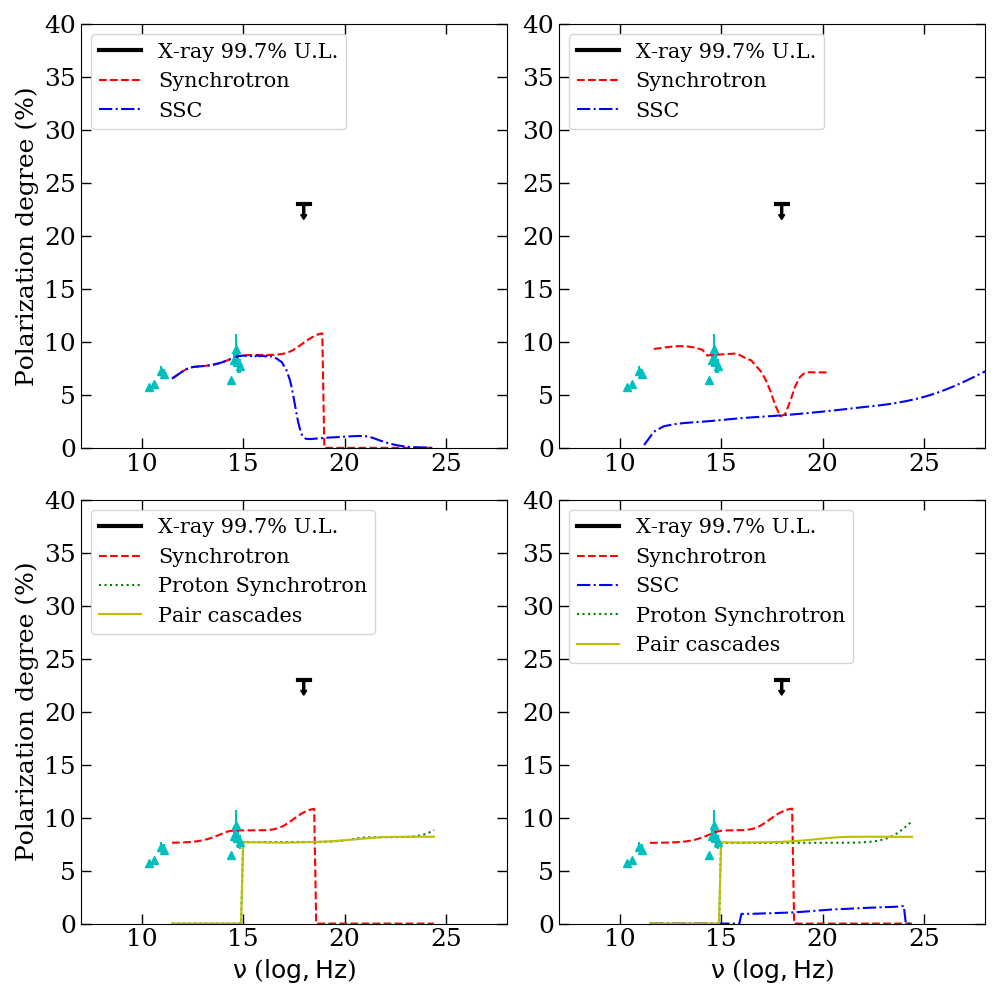}}
 \caption{Same as Fig. \ref{plt:SPD_OBS3_indi}, but for  SEG3. }
    \label{plt:SPD_SEG3_indi}
\end{figure}

\begin{figure}[H]
\centering
 \resizebox{\hsize}{!}{\includegraphics[width=\textwidth]{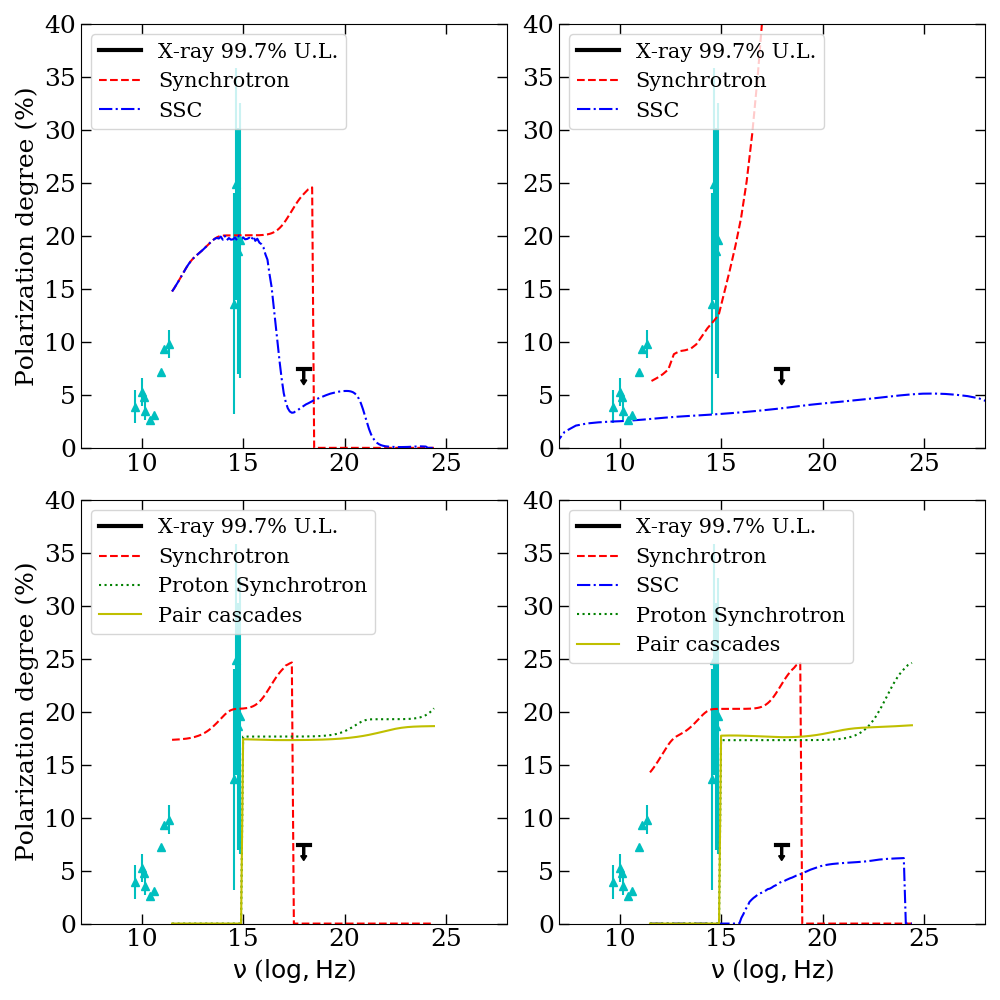}}
 \caption{Same as Fig. \ref{plt:SPD_OBS3_indi}, but for  OBS4. }
    \label{plt:SPD_OBS4_indi}
\end{figure}

\newpage

\section{Spectral energy distribution model parameters}
\begin{table*}
\centering
\caption{Parameter values for different observations for the pure SSC multi-zone model at the base of the jet.}
\label{tab:ssc}
\begin{tabular}{cccccc}\hline\hline

SSC & OBS3 & OBS3\_SEG1 & OBS3\_SEG2 & OBS3\_SEG3 & OBS4 \\ 
\hline
Bulk Lorentz factor $\Gamma$ & 10.3 & 10.3 & 10.3 & 10.3 & 3.037 \\\hline
Blob radius $R\, (\mathrm{cm})$ & $9.66 \times 10^{14}$ & $1.14 \times 10^{15}$ & $1.60 \times 10^{15}$ & $1.67 \times 10^{15}$ & $2.59 \times 10^{16}$ \\\hline
Magnetic field strength $B_0\, (\mathrm{G})$ & 0.4 & 0.25 & 0.3 & 0.3 & 0.1 \\\hline
Jet Power $W_j\, (\mathrm{erg/sec})$ & $1.7 \times 10^{45}$ & $1.5 \times 10^{45}$ & $2.8 \times 10^{45}$ & $3 \times 10^{45}$ & $6 \times 10^{45}$ \\\hline
Electron lower spectral cutoff $\gamma_{e,\min}$ & $7.94  $ & $8.12  $ & $8.12  $ & $8.12  $ & $1.02  $ \\\hline
Electron higher spectral cutoff$\gamma_{e,\max}$ & $1.58  \times 10^{4}$ & $2.51  \times 10^{4}$ & $1.40  \times 10^{4}$ & $1.40  \times 10^{4}$ & $1.00  \times 10^{4}$ \\\hline
Electron spectral index $\alpha$ & 1.46 & 1.76 & 1.4 & 1.46 & 1.3 \\\hline
Opening angle $\theta_{\mathrm{open}}$ & 12.77 & 12.77 & 12.77 & 12.77 & 10.06 \\\hline
Observing angle $\theta_{\mathrm{obs}}$ & 1.489 & 1.489 & 1.489 & 1.489 & 1.489 \\\hline
Equilibrium factor $A_{eq}$ & 0.8 & 0.4 & 0.7 & 0.7 & 2.68 \\\hline
 \# zones $N$& 20 & 27 & 39  & 40  & 22 \\\hline

\end{tabular}
\end{table*}

\begin{table*}
\centering
\caption{SSC+EC model parameters across all observations.}
\label{tab:ssc_ec}
\begin{tabular}{cccccc}\hline\hline

SSC+EC & OBS3 & SEG1 & SEG2 & SEG3 & OBS4 \\ 
\hline
Bulk Lorentz factor $\Gamma$ & 15 & 15 & 15 & 15 & 15 \\\hline
Blob radius $R\, (\mathrm{cm})$ & $8 \times 10^{15}$ & $5 \times 10^{15}$ & $1.2 \times 10^{16}$ & $1.2 \times 10^{16}$ & $2.2 \times 10^{16}$ \\\hline
Particle escaping timescale $\tau_{\mathrm{esc}}\, (\mathrm{s})$ & $1.0 \times 10^6$ & $6.7 \times 10^5$ & $1.6 \times 10^6$ & $1.6 \times 10^6$ & $3.0 \times 10^6$ \\\hline
Magnetic field strength $B\, (\mathrm{G})$ & 1.5 & 1.5 & 1.8 & 1.8 & 1.5 \\\hline
Electron injection luminosity $L_e\, (\mathrm{erg/s})$ & $1.0 \times 10^{45}$ & $1.0 \times 10^{45}$ & $1.0 \times 10^{45}$ & $1.2 \times 10^{45}$ & $4.0 \times 10^{44}$ \\\hline
Electron lower spectral cutoff $\gamma_{e,\min}$ & $2 \times 10^3$ & $2 \times 10^3$ & $1.5 \times 10^3$ & $1.5 \times 10^3$ & $4.5 \times 10^2$ \\\hline
Electron higher spectral cutoff $\gamma_{e,\max}$ & $1 \times 10^5$ & $1 \times 10^5$ & $1 \times 10^5$ & $5 \times 10^4$ & $3 \times 10^4$ \\\hline
Electron spectral index $p_e$ & 3.0 & 2.9 & 3.2 & 3.2 & 3.4 \\\hline
External photon field temperature $T\, (\mathrm{K})$ & $6 \times 10^3$ & $6 \times 10^3$ & $6 \times 10^3$ & $6 \times 10^3$ & $6 \times 10^3$ \\\hline
External photon energy density $u_{\mathrm{ext}}\, (\mathrm{erg/cm^3})$ & $2 \times 10^{-4}$ & $2 \times 10^{-4}$ & $2 \times 10^{-4}$ & $2 \times 10^{-4}$ & $2 \times 10^{-5}$ \\\hline

\end{tabular}
\end{table*}

\begin{table*}
\centering
\caption{SSC+hadronic model parameters across all observations.}
\label{tab:ssc_hadronic}
\begin{tabular}{cccccc}\hline\hline

SSC+hadronic & OBS3 & SEG1 & SEG2 & SEG3 & OBS4 \\ 
\hline
Bulk Lorentz factor $\Gamma$ & 15 & 15 & 15 & 15 & 15 \\\hline
Blob radius $R\, (\mathrm{cm})$ & $1.1 \times 10^{15}$ & $8 \times 10^{14}$ & $1.3 \times 10^{15}$ & $1.2 \times 10^{15}$ & $1.1 \times 10^{15}$ \\\hline
Particle escaping timescale $\tau_{\mathrm{esc}}\, (\mathrm{s})$ & $1.5 \times 10^5$ & $1.0 \times 10^5$ & $1.7 \times 10^5$ & $1.6 \times 10^5$ & $1.5 \times 10^5$ \\\hline
Magnetic field strength $B\, (\mathrm{G})$ & 50 & 50 & 50 & 50 & 10 \\\hline
Electron injection luminosity $L_e\, (\mathrm{erg/s})$ & $2.8 \times 10^{44}$ & $2.0 \times 10^{44}$ & $3.0 \times 10^{44}$ & $3.5 \times 10^{44}$ & $9.5 \times 10^{43}$ \\\hline
Electron lower spectral cutoff $\gamma_{e,\min}$ & $3 \times 10^2$ & $3 \times 10^2$ & $3 \times 10^2$ & $3 \times 10^2$ & $3.5 \times 10^2$ \\\hline
Electron higher spectral cutoff $\gamma_{e,\max}$ & $1 \times 10^5$ & $1 \times 10^5$ & $1 \times 10^5$ & $5 \times 10^3$ & $2 \times 10^4$ \\\hline
Electron spectral index $p_e$ & 3.3 & 3.3 & 3.3 & 3.3 & 3.7 \\\hline
Proton injection luminosity $L_p\, (\mathrm{erg/s})$ & $5.3 \times 10^{46}$ & $6.5 \times 10^{46}$ & $6.0 \times 10^{46}$ & $5.5 \times 10^{46}$ & $2.0 \times 10^{47}$ \\\hline
Proton higher spectral cutoff $\gamma_{p,\max}$ & $8 \times 10^8$ & $8 \times 10^8$ & $8 \times 10^8$ & $8 \times 10^8$ & $5 \times 10^8$ \\\hline

\end{tabular}
\end{table*}

\begin{table*}
\centering
\caption{Hadronic model parameters across all observations.}
\label{tab:hadronic}
\begin{tabular}{cccccc}\hline\hline

Hadronic & OBS3 & SEG1 & SEG2 & SEG3 & OBS4 \\ 
\hline
Bulk Lorentz factor $\Gamma$ & 15 & 15 & 15 & 15 & 15 \\\hline
Blob radius $R\, (\mathrm{cm})$ & $8 \times 10^{15}$ & $8 \times 10^{15}$ & $8 \times 10^{15}$ & $8 \times 10^{15}$ & $8 \times 10^{15}$ \\\hline
Particle escaping timescale $\tau_{\mathrm{esc}}\, (\mathrm{s})$ & $1.0 \times 10^6$ & $1.0 \times 10^6$ & $1.0 \times 10^6$ & $1.0 \times 10^6$ & $1.0 \times 10^6$ \\\hline
Magnetic field strength $B\, (\mathrm{G})$ & 30 & 30 & 30 & 30 & 30 \\\hline
Electron injection luminosity $L_e\, (\mathrm{erg/s})$ & $3.0 \times 10^{44}$ & $3.0 \times 10^{44}$ & $3.0 \times 10^{44}$ & $4.0 \times 10^{44}$ & $6.5 \times 10^{43}$ \\\hline
Electron lower spectral cutoff $\gamma_{e,\min}$ & $2 \times 10^2$ & $2 \times 10^2$ & $2 \times 10^2$ & $2 \times 10^2$ & $2 \times 10^2$ \\\hline
Electron higher spectral cutoff $\gamma_{e,\max}$ & $2 \times 10^4$ & $2 \times 10^4$ & $2 \times 10^4$ & $7 \times 10^3$ & $2 \times 10^3$ \\\hline
Electron spectral index $p_e$ & 3.3 & 3.7 & 3.3 & 3.3 & 3.7 \\\hline
Proton injection luminosity $L_p\, (\mathrm{erg/s})$ & $5.5 \times 10^{47}$ & $5.5 \times 10^{47}$ & $5.5 \times 10^{47}$ & $5.5 \times 10^{47}$ & $5.5 \times 10^{47}$ \\\hline
Proton break energy $\gamma_{p,br}$ & $5 \times 10^5$ & $5 \times 10^5$ & $5 \times 10^5$ & $5 \times 10^5$ & $5 \times 10^5$ \\\hline

\end{tabular}
\end{table*}

\end{appendix}

\end{document}